\pdfoutput=1
\documentclass[sigconf,natbib=true]{acmart}

\usepackage{enumitem}
\usepackage{booktabs}
\usepackage{xspace}

\newcommand{\an}[1]{$^{_{_{^{^{#1}}}}}$}

\setcopyright{none}
\renewcommand\footnotetextcopyrightpermission[1]{}

\begin{document}

\title{Can Old TREC Collections Reliably Evaluate\\ Modern Neural Retrieval Models?}

\author{Ellen M. Voorhees,\an{1} Ian Soboroff,\an{1} Jimmy Lin\an{2}}

\affiliation{\vspace{0.1cm}
$^1$ National Institute of Standards and Technology, Gaithersburg, Maryland \country{USA} \\
$^2$ David R. Cheriton School of Computer Science, University of Waterloo, Ontario \country{Canada} \\
}

\renewcommand{\shortauthors}{}
\pagestyle{empty}

\begin{abstract}
Neural retrieval models are generally regarded as fundamentally different from the retrieval techniques used in the late 1990's when the TREC ad~hoc test collections were constructed.
They thus provide the opportunity to empirically test the claim that pooling-built test collections can reliably evaluate retrieval systems that did not contribute to the construction of the collection (in other words, that such collections can be {\em reusable}).
To test the reusability claim, we asked TREC assessors to judge new pools created from new search results for the \mbox{TREC-8} ad~hoc collection.
These new search results consisted of five new runs (one each from three transformer-based models and two baseline runs that use BM25) plus the set of \mbox{TREC-8} submissions that did not previously contribute to pools.
The new runs did retrieve previously unseen documents, but the vast majority of those documents were not relevant.
The ranking of all runs by mean evaluation score when evaluated using the official \mbox{TREC-8} relevance judgment set and the newly expanded relevance set are almost identical, with Kendall's $\tau$ correlations greater than $0.99$.
Correlations for individual topics are also high.
The \mbox{TREC-8} ad~hoc collection was originally constructed using deep pools over a diverse set of runs, including several effective manual runs.
Its judgment budget, and hence construction cost, was relatively large.
However, it does appear that the expense was well-spent:\ even with the advent of neural techniques, the collection has stood the test of time and remains a reliable evaluation instrument as retrieval techniques have advanced.
\end{abstract}

\maketitle

\section{Introduction}

A primary motivation for the Text REtrieval Conferences (TRECs) is to build large test collections for the information retrieval research community~\cite{TRECintro}.
The goal is for these collections to be reusable---meaning that retrieval systems that did not participate in the collection-building process could still be evaluated fairly using them---and, in particular, to be useful for evaluating systems that did not exist at the time the collection was built.
Examination of the TREC ad~hoc collections shortly after they were built supported the conclusion that the collections are indeed reusable~\cite{TRECcolls,zob:sigir98}.
The \mbox{TREC-8} ad~hoc collection has been considered an especially reliable collection given both the high-quality (manual) runs that contributed to its construction and the large set of relevance judgments made for it~\cite{CIKMbandits}.

Parts of the community, however, have been skeptical of the viability of using TREC ad~hoc collections to research new neural retrieval models.
The primary argument has been that since neural methods ``work very differently'' from ``traditional'' retrieval models, they would retrieve many previously unjudged documents that are relevant.
Neural methods would not be properly rewarded for actually ``being better'', and thus it would be misleading to assess progress based on these incomplete evaluation instruments.
This line of reasoning is, for example, explicitly stated in Yilmaz et al.~\cite{Yilmaz_etal_SIGIR2020}, and the TREC Common Core track was started in large part so that the newer retrieval models could contribute to the construction of additional ad~hoc collections.\footnote{See \url{https://trec-core.github.io/2018/}.}

The \mbox{TREC-8} collection, the last of the TREC ad~hoc collections, was created in 1999, long before the emergence of the current neural models.
We can thus use these models to test the reusability claims of the original TREC ad~hoc collections.
This paper reports on one such test, a look at how the \mbox{TREC-8} ad~hoc collection evaluates representative runs from three trans\-former-based retrieval models (a reranker, a dense retrieval model, and a sparse retrieval model) and two new baselines (BM25-based). 
New pools created from these five runs and a set of \mbox{TREC-8} submissions that did not previously contribute to the pools were judged by TREC assessors.
Some new relevant documents were found, as expected, but most of the newly retrieved (and previously unjudged) documents were judged {\it not} relevant.
The ranking of systems by mean score when evaluated using the official \mbox{TREC-8} relevance judgment set and the newly expanded relevance judgment set are almost identical with Kendall's~$\tau$ correlations greater than 0.99.
Correlations for individual topics are also high.

Thus, the answer to the question posed in the title appears to be, {\it yes}, at least for the TREC-8 collection examined in our experiments.
The contribution of this paper is, to our knowledge, the first time this question has be rigorously tackled and answered.
While there are additional nuances to this high-level finding (see Section~\ref{section:discussion}), it does appear that this well-built test collection has stood the test of time and remains a reliable evaluation instrument, even as retrieval techniques have advanced significantly.

\section{Background and Related Work}

A retrieval test collection consists of a set of documents, a set of information needs called {\em topics} that can be met by those documents, and a set of relevance judgments that say which documents should be retrieved for which topics.
The set of judgments in a collection is often referred to as the {\em qrels} (short for query-relevance), a convention we will follow in this paper.
Given a test collection, the retrieval output of a search engine (a ranked list of documents retrieved for each topic and called a {\em run}) can be evaluated using a variety of measures that are functions of the ranks at which relevant documents are retrieved. 

The very first retrieval test collections had {\em complete} judgments; that is, every document was judged by a human for every topic.
However, complete judgments are only practical for small test collections, and small test collections are not representative of the challenges operational search systems encounter.
To build larger test collections, some sort of sampling procedure is needed so that for each topic a human judge looks at only a tiny portion of the entire document set.

TREC was the first to implement a process called {\em pooling}~\cite{pooling} to sample the document corpus and build much larger test collections than were previously available.
In pooling, the set of documents to be judged for a topic, the {\em pool}, is the union of the documents retrieved in the top $\lambda$ ranks over a given set of runs (such as the runs submitted to a particular evaluation, for example).
Larger values of $\lambda$ lead to more documents in the pools and produce {\em deeper} pools than smaller values of $\lambda$.
The assessor for the topic assigns a relevance judgment to each document in the pool.
Any document that was not in the pool is thus not judged; any such unjudged document encountered in the evaluation of a run is assumed to be not relevant.
The rationale of pooling is the belief that taking sufficiently many top-ranked documents from a diverse set of effective runs will capture most relevant documents such that treating all other documents as not relevant can still yield a reliable evaluation.

All of the early TREC collections, including the \mbox{TREC-8} ad~hoc collection, were built using pooling.
Every year, each TREC participant submitted a handful of different runs.
Pools were constructed using a subset of the submitted runs from each participant, with the total number of runs contributing to the pools determined by the judgment budget.
Runs that contributed to the pools are called ``judged runs'' and the remainder are ``unjudged runs''.
Furthermore, each run is also designated as being either automatic or manual.
An automatic run is a run in which there was no manual intervention of any kind to produce the ranked lists of documents from the topic statements; a manual run is anything else, which may encompass simple tweaks to the topic statement to intensive interaction with a retrieval system (including manual and possibly iterative query formulation, relevance assessment for feedback, etc.).

Zobel showed that the quality of a collection built through pooling depends on both the diversity of the runs and the depth ($\lambda$) to which the pools were constructed, but found the TREC collections of the day to be reliable in that they evaluated unjudged runs fairly~\cite{zob:sigir98}.
Analysis of the \mbox{TREC-8} collection immediately after its construction using a variant of the process Zobel used found it, too, to be reliable~\cite{TREC8over}.
The process simulated the evaluation of ``new'' retrieval methods by removing from the qrels those relevant documents that only a single participant contributed to the pools and comparing the evaluation of that participant's runs when using either the full or the reduced qrels.
For \mbox{TREC-8}, manual runs contributed most of the unique relevant documents and were consequently most affected by the removal of their uniquely retrieved relevant documents.
The change in evaluation scores for TREC-8 automatic runs with and without their own uniques was negligible, probably because the manual runs had retrieved so many relevant documents.
The quality of the pools is known to be significantly enhanced by the presence of recall-oriented manual runs such that the organizers of the NTCIR workshops performed their own manual runs to supplement the pools when building their first collections~\cite{NTCIR}.

Unfortunately, pooling has its own size dependency and cannot be used to create reliable collections for arbitrarily large document corpora without arbitrarily large judgment budgets~\cite{IRpooling}.
New ways of building test collections and new evaluation measures that accommodate missing judgments continue to be active research areas.
The available tools for gauging test collection quality, such as the uniques test, all rely on runs available during collection construction, and thus indicate true problems when they detect a problem with a collection but may not detect problems that nonetheless exist (the ``unknown unknowns'').
This paper does not offer new methods for gauging collection quality, but our results do confirm that the \mbox{TREC-8} collection scored our new neural runs fairly.

\section{Methods}

This section first describes the process used to obtain new judgments for the \mbox{TREC-8} collection and then describes the new runs themselves.

\subsection{Pooling}
The \mbox{TREC-8} ad~hoc collection contains approximately 525,000 full-text documents drawn from the {\em Financial Times}, the {\em Los Angeles Times}, the {\em Foreign Broadcast Information Service}, and the {\em Federal Register} and 50 topics (numbers 401--450).
The TREC task from which the collection was created received a total of 129 runs from 41 participants.
Pools were created using $\lambda=100$ over 71 of these runs resulting in a total of 86,830 judged documents across all 50 topics with the smallest pool containing 1046 documents and the largest 2992.
Thirteen of the submitted runs were manual runs and the rest were automatic runs.
See the \mbox{TREC-8} overview paper for more details about the collection~\cite{TREC8over}.

The five new runs that are the subject of this analysis consist of two BM25-based baselines and representative samples of three transformer-based retrieval models.  These five runs are described in detail in Section~\ref{section:newruns}.
For the collection to be unfair to these runs, the runs would have to retrieve unjudged documents that are in fact relevant, and determining that requires additional human relevance judgments.
But adding new judgments to an existing test collection is fraught with complications.
Relevance is known to be idiosyncratic to the individual assessor making the judgment~\cite{IPMrel} and cherry-picking documents from a small set
of runs risks biasing the qrels in favor of those runs.
We used the following procedure to obtain new judgments to control for these factors as much as possible.

We constructed depth-50 pools using the five new runs and 52 of the 58 unjudged runs submitted to \mbox{TREC-8}.
The six unjudged runs from \mbox{TREC-8} that were again not pooled are ineffective runs (MAP scores less than 0.1 as evaluated on the original qrels) that contain disproportionately many unjudged documents.
For each topic, any previously-judged document in the new pool was removed, and
a TREC assessor judged the remainder.
In keeping with the original TREC-8 judgment protocol, assessors assigned binary judgments of not relevant or relevant to each document in the (remainder) pool, and were instructed to judge a document as relevant if any part of it was relevant.

The TREC assessor for a topic for the new pools was not the same assessor for that topic as in \mbox{TREC-8}.
(The \mbox{TREC-8} assessors were not available, and after 20+ years since they last assessed the topics they would have been essentially different assessors, anyway.)
The current assessor was given access to the previous qrels and asked to review those judgments to get a sense of how the original assessor judged the topic before beginning their own judgments.
The combined set of judgments may well be less internally-consistent than the original set, but any such conflicts are unlikely to matter for this experiment.
The TREC assessor has no knowledge of which system retrieved which documents and so cannot be systematically biased for or against particular systems.
Historically, assessors like to find relevant documents so they are unlikely to arbitrarily declare documents to be not relevant.
Further, assessors tend to disagree on ``edge case'' documents and our main concern is new runs retrieving clearly relevant but previously unseen documents.

\begin{figure}
\includegraphics[width=\columnwidth]{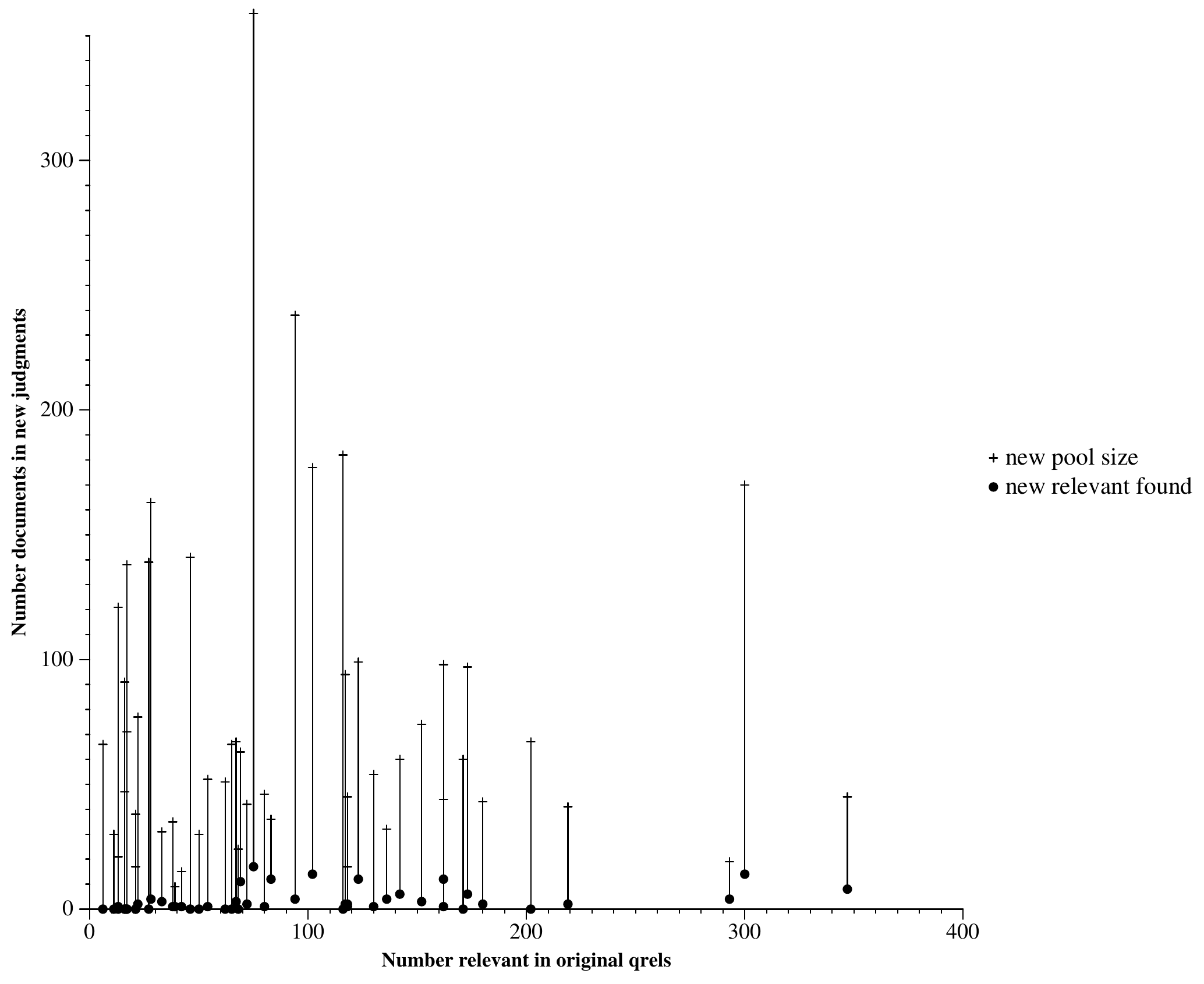}
\caption{The number of documents in the remainder pool and the number relevant documents found in it conditioned on the number of relevant documents in the original qrels for each of the 50 topics. No new relevant documents were found for 18 of the topics.}
\label{counts.fig}
\end{figure}

The total number of documents in the remainder pools was 3842 with the smallest pool containing 9 documents and the largest 359.
The total number of relevant documents found is 158, with 17/359 new relevant documents found for topic 417 and no new relevant documents found for 18 topics.
Figure~\ref{counts.fig} shows the number of newly judged documents and the corresponding number of relevant documents found per topic (on the $y$-axis) conditioned on the number of relevant documents for the topic in the original qrels (on the $x$-axis).
Contrary to Zobel's~\cite{zob:sigir98} and Harman's~\cite{TRECcolls} findings that topics with large relevant set sizes have even more relevant in the unjudged documents, no such correlation between the number of existing and newly found relevant documents is apparent in this case.
Consistent with their findings, though, the newly found relevant documents are not concentrated in a small set of runs (see Figure~\ref{unj-rel.fig}).

\subsection{Retrieval Runs}
\label{section:newruns}

We began with two bag-of-words baselines produced by the Anserini IR toolkit, which is built on the open-source Lucene search library to support reproducible research~\cite{Yang_etal_JDIQ2018}:

\begin{itemize}[leftmargin=*]

\item Anserini BM25:\ Lucene's implementation of BM25~\cite{Robertson_Zaragoza_FnTIR2009}, which can be viewed as a BM25 variant (see detailed discussions in Kamphuis et al.~\cite{Kamphuis_etal_ECIR2020}).

\item Anserini BM25+RM3: BM25 with the RM3~\cite{Abdul-Jaleel04} pseudo relevance feedback technique, as described in Yang et al.~\cite{Yang_etal_SIGIR2019}.
This provides a competitive baseline, especially with respect to pre-BERT neural models.

\end{itemize}

\noindent In addition, we also generated three new runs with neural models:

\begin{itemize}[leftmargin=*]

\item monoBERT + MaxP: a reranking model based on monoBERT~\cite{Nogueira:1901.04085:2019} that takes advantage of the MaxP technique~\cite{Dai_Callan_SIGIR2019} to overcome the length limitations associated with transformers.
Here, we rerank the output of BM25 from Anserini (see above).
Our implementation is described in Zhang et al.~\cite{ZhangXinyu_etal_ECIR2021} and trained on the MS MARCO (V1) passage data.\footnote{https://github.com/microsoft/msmarco}

\item TCT-ColBERT (v2)~\cite{Lin_etal_2021_RepL4NLP}:\ a representative example of the class of so-called dense retrieval models that takes advantage of transformers to convert documents into dense vectors.
Retrieval is then recast as a nearest neighbor search problem in vector space.
To address the length limitations associated with transformers, documents are first segmented into passages, and each passage is encoded independently.
At retrieval time, the highest-scoring passage score is taken as the score of the document it came from to generate a document ranking for evaluation.
The encoder models are trained on the MS MARCO (V1) passage data.

\item uniCOIL (with doc2query--T5 expansions)~\cite{Lin_Ma_arXiv2021}: a representative example of the class of so-called sparse retrieval models.
These models likewise take advantage of transformers to generate vector representations from documents and queries, but the main difference here is that these models retain the vocabulary space as the basis of the vectors, and thus they can be viewed as bag-of-words weighting functions that are {\it learned} from large amounts of data.
As with TCT-ColBERT, documents are segmented into passages and independently encoded, and retrieval (which can be performed with standard inverted indexes) likewise takes the highest passage score as the document score.
The encoder models are trained on the MS MARCO (V1) passage data.

\end{itemize}

\noindent Together, these models cover the three main ways that transformers are used today for retrieval~\cite{Lin_etal_2021_ptr4tr}:\ reranking bag-of-words candidates, dense retrieval models, and sparse retrieval models.
At a high level, while none of the three would be considered ``state of the art'' (SOTA) in terms of standard benchmark datasets, they can be fairly characterized as competitive models against which putative SOTA models would be evaluated.

Note that for simplicity, these models have all been trained on the MS MARCO (V1) passage test collection and applied for retrieval (inference)  in a zero-shot manner.
For reranking approaches (e.g., monoBERT), there is substantial evidence that they are able to maintain high levels of effectiveness even when applied to texts beyond the domain on which it is trained~\cite{Yilmaz_etal_EMNLP2019,Li:2008.09093:2020,Lin_etal_2021_ptr4tr}.
That is, rerankers exhibit good cross-domain generalizations with respect to relevance.
In contrast, there is evidence that dense retrieval models in general have difficulty with cross-domain generalization, with evidence from multi-domain datasets such as BIER~\cite{Thakur:2104.08663:2021}.
There appears to be some evidence that sparse retrieval models may generalize across domains better~\cite{Formal:2109.10086:2021}, but evidence here is more scant.
The cross-domain generalization deficiencies of dense and sparse retrieval models (and how to rectify the situation) is the subject of ongoing research, but to our knowledge there have not emerged best practices that we can simply ``drop in'' for these experiments.
Thus, we decided on zero-shot inference so as to not conflate aspects of modeling approaches not germane to our research question.
We acknowledge that this is a weakness in our design, as we further discuss in Section~\ref{section:discussion}.

\section{Results}
\label{section:eval}

Our research question is whether the five new runs described in Section~\ref{section:newruns} are evaluated fairly by the original \mbox{TREC-8} test collection.
Put differently, would a researcher using the original test collection to compare the effectiveness of one of the new runs to a \mbox{TREC-8} submission (that contributed to the pools) reach the same conclusion had the new run also contributed to the pools?
To answer this question, we simply evaluate all 134 runs (129 TREC-8 submissions plus 5 new runs) using both the original qrels and an expanded qrels that is the union of the original plus new judgments and rank the runs by mean evaluation score.
If the two rankings of runs are almost the same, this suggests that the new runs can indeed be fairly evaluated, and that the original collection is reliable.

We use Kendall's $\tau$ measure of association~\cite{Kendall} as the similarity measure of system rankings to operationalize ``almost the same''.
Kendall's $\tau$ computes a normalized count of the number of pairwise swaps it takes to turn one ranking into the other. The $\tau$ ranges from $-1.0$ to $1.0$ where $1.0$ indicates the rankings are identical, $-1.0$ indicates the rankings are exact opposites of one another, and $0.0$ indicates the rankings are uncorrelated.
Kendall's $\tau$ is not an ideal similarity measure~\cite{tauProblems}.
Its values depend on the number of items being ranked, so are granular when there are few items.
The values and are also sensitive to the average difference in mean scores, so small differences in average scores that are not meaningful in practice may still change the order of systems making rankings look less similar than they actually are.
But for our purposes where there are 134 runs in the ranking and the rankings are generally stable $\tau$ is less problematic.
The implementation of Kendall's $\tau$ used here handles tied scores in the rankings by omitting the tied run pair from the computation.

We used both mean average precision (MAP) and mean precision at ten documents retrieved (P@10) as evaluation measures. 
The Kendall's $\tau$ between system rankings for MAP is $0.9933$ and for P@10 is $0.9991$, indicating very consistent ranking of systems by the two different qrels.
Such small differences in rankings are well within the noise level of information retrieval evaluation, for example, changes in relevance assessors create larger differences~\cite{IPMrel}.

Table~\ref{map.tab} reports the MAP scores and corresponding ranks over all evaluated runs for the best run overall (a manual run), the best automatic run, the median run, and the five new runs as computed using the original and expanded qrels.
As the $\tau$ indicates, the ranks of the runs change minimally.
The absolute value of the MAP scores decreases when computed using the expanded qrels:\ the expanded recall base decreases the runs' scores more than retrieving additional relevant documents helps since each run retrieves at most only a few additional relevant documents.

Qualitatively, the effectiveness of the runs and their rank positions are generally within expectations, although there are a few surprises.
The bag-of-words BM25 baseline is roughly ``middle of the pack'', which makes sense since BM25 ``of today'' is likely not very different from BM25 and comparable bag-of-words models from two decades ago in terms of effectiveness.
Pseudo relevance feedback (RM3) improves over bag-of-words BM25, once again as expected, and the improvements are consistent with the literature.
It was also expected that we see improvements from monoBERT + MaxP compared to BM25 (which the reranker uses as a source of candidates), but the amount of improvement is somewhat disappointing---transformer-based reranking only achieves effectiveness comparable to pseudo relevance feedback.
That is, a lot of ``effort'' with neural models was expended to achieve only what can be obtained with a technique that is over a decade old.\footnote{We did not experiment with reranking BM25 + RM3.}
This result appears inconsistent with work on the TREC 2004 Robust Track (which uses the same document corpus as \mbox{TREC-8} but with additional topics), where researchers have reported quite impressive scores, even besting the most effective run from the participants~\cite{Yilmaz_etal_EMNLP2019,Li:2008.09093:2020}.
The fact that TCT-ColBERT and uniCOIL underperform the BM25 baseline is consistent with previous work, given that the model is trained on another collection and applied in a zero-shot manner~\cite{Thakur:2104.08663:2021}.
It is worth noting that in our experiments, these neural runs are the first and only runs that we generated---with no tuning (of, for example, inference-time hyperparameters) or consultation of the evaluation scores whatsoever.

\begin{table}[t]
\centering\scalebox{0.87}{
\begin{tabular}{lrrrr}
\toprule
Run & MAP (orig) & rank & MAP (exp) & rank \\
\toprule
Top manual run & 0.4692 & 1 & 0.4587 & 1 \\
Top automatic run & 0.3303 & 11 & 0.3262 & 11\\
median run & 0.2602 & 66 & 0.2568 & 67 \\
\midrule
BM25 & 0.2515 & 76 & 0.2497 & 74 \\
BM25 + RM3 & 0.2750 & 50 & 0.2721 & 50 \\
\midrule
monoBERT + MaxP & 0.2728 & 52 & 0.2721 & 51 \\
TCT-ColBERT & 0.2209 & 96 & 0.2198 & 96 \\
uniCOIL & 0.2343 & 85 & 0.2325 & 84 \\
\bottomrule
\end{tabular}}
\vspace{0.2cm}
\caption{The effectiveness of a few selected runs and our additional runs, using the original qrels (orig) and the expanded qrels (exp).
Ranks are out of 134 total runs (129 TREC-8 submissions plus 5 new runs).}
\label{map.tab}
\end{table}

\begin{figure}
\includegraphics[width=\columnwidth]{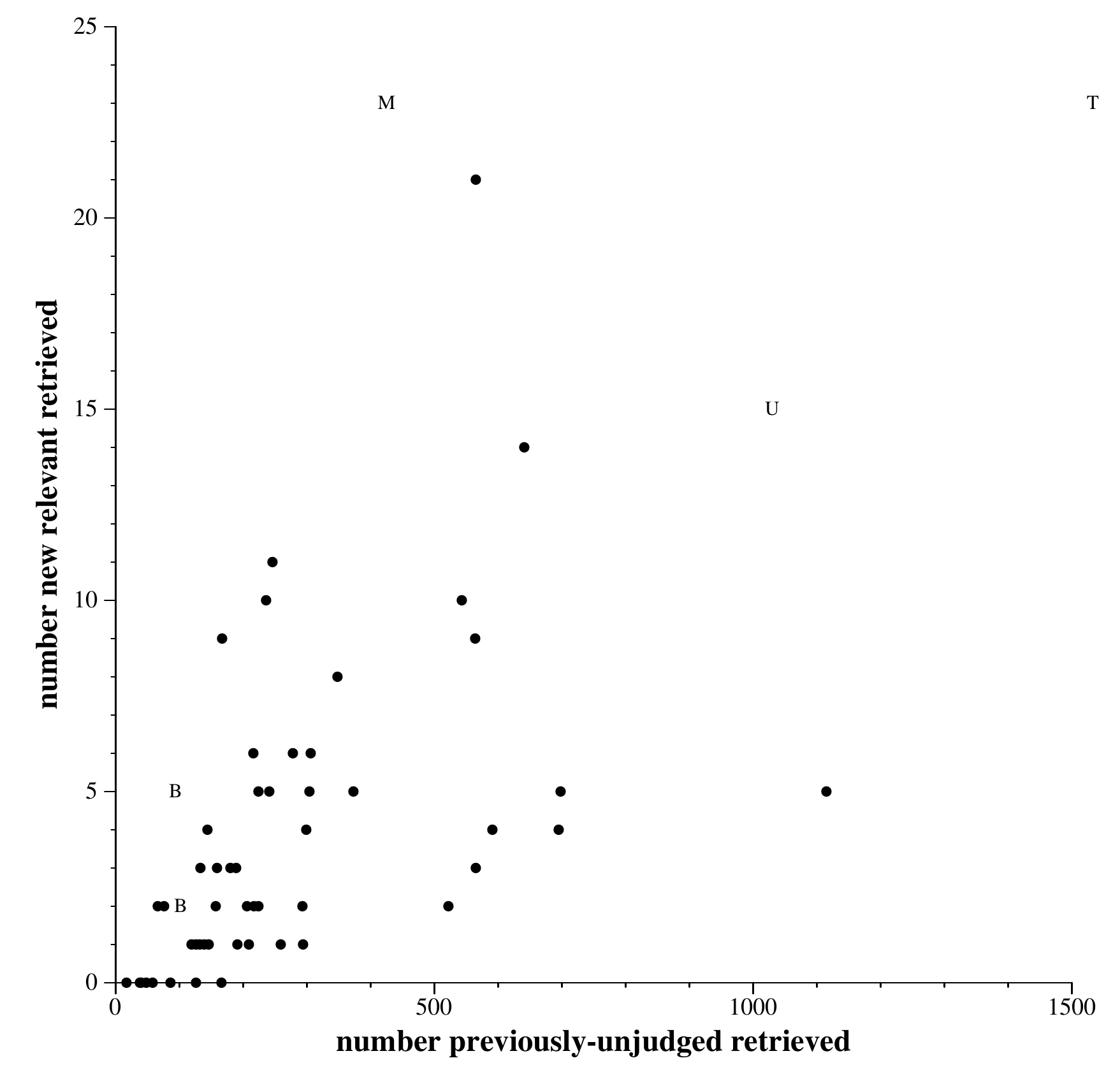}
\caption{Number of newly found relevant documents vs. number of previously unjudged documents found in the top 100 ranks totaled over all topics.  Unjudged TREC-8 submissions are plotted with a dot; new runs are plotted with the first character of their run name (B for both BM25 baselines, U for uniCOIL, T for TCT-ColBERT, and M for monoBERT). }
\label{unj-rel.fig}
\end{figure}

Examination of the number of newly found relevant documents per run, shown in Figure~\ref{unj-rel.fig}, explains why the rankings are as consistent as they are.
The figure plots the number of newly found relevant documents against the number of previously-unjudged documents retrieved by a run in the top 100 ranks over all topics for each of the 57 runs that contributed to the new pools.
The TCT-ColBERT run returned both more previously unjudged documents and more newly found relevant documents than any of the \mbox{TREC-8} submissions, and
monoBERT+maxP and TCT-ColBERT each retrieved the maximum of 23 newly found relevant documents.
But 23 additional relevant documents in a run is an average of slightly less than one additional relevant document for every two topics, and most runs added fewer than 10 additional relevant (an average of one additional relevant for every five topics).

Since averages can often hide significant variance among individual topics,
we computed per-topic $\tau$ scores to check for topics that were impacted by newly found relevant documents.
For each of the 50 topics, we ranked the systems by their scores on that topic as computed using each of the qrels and computed the $\tau$ between the two rankings.
For P@10, topic 401 had a $\tau$ of $0.9988$ and all other topics had a $\tau$ of $1.0$.
There was more variability for MAP with individual topic $\tau$ values ranging from $0.8852$ to $1.0$ (the 18 topics with no newly found relevant had $\tau$ values of $1.0$).
The one topic with $\tau < 0.9$ is topic 432 that has 28 relevant documents in the original qrels and an additional four relevant documents were found for it.
Most automatic runs have very poor effectiveness for topic 432; the several runs that retrieved one or two of the newly found relevant documents had very large changes in rank even though the average precision score did not change much in absolute terms.
The run with the largest change in ranks was the monoBERT run that improved 53 ranks with a change in AP score from $0.0022$ to $0.0188$.

\section{Discussion}
\label{section:discussion}

The ranking of runs by the expanded qrels is nearly identical to the one when ranked by the original qrels, so we conclude that the \mbox{TREC-8} collection is reusable.
This conclusion appears to contradict the warning by Yilmaz et al.\ that the evaluation of neural methods on collections built solely from traditional methods may be unfair~\cite{Yilmaz_etal_SIGIR2020}, but the source of disagreement is in the quality of the respective collections.
The quality of a pooled collection is strongly dependent on the effectiveness of the runs from which the pools were taken, the diversity of those runs, and the pool depth~\cite{zob:sigir98,philosophy}.
In contrast, Yilmaz et al.\ were explicitly interested in shallow pools, and so the resulting evaluation using them can be expected to be noisy.
Our claim of reusability is specific to the \mbox{TREC-8} collection, though the results are likely to apply to other similarly-constructed collections such as the \mbox{TREC-6} and \mbox{TREC-7} ad~hoc collections.

Our experiments also do not prove that there cannot be some other retrieval method that would produce a run---we'll call {\em magical}---that would in fact be evaluated unfairly.
The only way to prove that is to judge the entire document set for each topic.\footnote{Including the new judgments, on average each topic has 1800 judged documents (or $0.3$\% of the document corpus).}
However, we believe that the existence of run {\em magical} is unlikely.
For {\em magical} to be evaluated unfairly it would have to both discover a sufficient number of new relevant documents and also rank those new relevant before the majority of the known relevant for sufficiently many topics (otherwise, the existing qrels will score it correctly).
Such a result is made even more unlikely by the fact that TREC topic development process, in which the TREC assessor who authors a topic performs a few manual searches to estimate the likely number of relevant, was designed to create topics that are expected to have fairly limited relevant sets~\cite{TRECcolls}. 

The argument against the existence of run {\em magical} also addresses a possible objection to our experimental methodology:\ that the dense retrieval model (TCT-ColBERT) and the sparse retrieval model (uniCOIL) were used in a zero-shot manner.
Due to challenges in cross-domain inference, the effectiveness of both approaches is low, even worse than bag-of-words BM25 with Anserini (which is unsurprising).
Had we performed appropriate domain adaptation on the dense and sparse retrieval models, they would have been more effective overall, and perhaps would have uncovered more previously unseen but relevant documents.
As already argued, such a result is possible, but unlikely.
Furthermore, there has not, to our knowledge, emerged a consensus on domain adaptation best practices for such models, and thus any technique we apply runs the risk of being idiosyncratic.
To appropriately use in-domain training data would necessitate some type of cross-validation setup, which would render our experimental setup needlessly complicated.
We feel that we have made the appropriate design choices in this first attempt to answer our core research question, and leave more nuanced examination of these additional factors for future work.

The quality of the \mbox{TREC-8} collection has long been attributed to the effective manual runs that contributed to the pools during its construction.
In principle, it is the effectiveness of a run and not the type of the method that creates it that matters, but to date only manual enrichment of the pools has been sufficiently effective.
Since neural methods, especially dense retrieval models, attempt to overcome the kinds of semantic mismatch that traditional bag-of-words methods are susceptible to and humans handle with ease, we examined whether the original TREC-8 collection would have been as good had it been built using the original \mbox{TREC-8} automatic runs and the new neural runs but no manual runs.

To accomplish this, we removed all the relevant documents that had been contributed to the original pools by manual runs only, and merged those modified pools with depth-100 pools built from the five new runs and the originally unjudged \mbox{TREC-8} submissions (minus the three manual submissions from that set).
Of the 1131 relevant documents that only manual runs contributed to the original pools across all 50 topics, fewer than 100 were recovered by the new pools.
The TCT-ColBERT run found 50 of these documents, the monoBERT run found 22, and the uniCOIL run 17 (some of the same documents were found by more than one run).
Each of the remaining runs found substantially fewer documents; the BM25+RM3 run found just two, for example.
So there is some support for the claim that the neural methods are finding different documents than traditional (automatic) methods.

There is a total of 4728 relevant documents in the original qrels, so the loss of 1000 relevant documents is almost a fifth of the known relevant.
Nonetheless, the Kendall's $\tau$ correlation between rankings of runs evaluated with the original qrels and evaluated with the minus-manual-runs qrels is still very high ($0.9964$ for P@10 and $0.9818$ for MAP).
How can this be so?
The manual runs are so much better than the other runs that they still evaluate as better using only the documents they retrieved in common with the automatic runs;
the automatic runs did not retrieve the lost documents in the top 100 ranks (by definition of them being manual-only), so were mostly unaffected by their loss; and there are only three neural runs (which did have modest movement in the system rankings).

\section{Conclusion}

The \mbox{TREC-8} ad~hoc collection was originally constructed in 1999 using deep pools over a diverse set of runs including several effective manual runs that retrieved many relevant documents not retrieved by other runs.
Containing 86,830 judgments over 50 topics, the collection was one of the most expensive TREC collections to build, an expense justified by the belief that the resulting collection would be reusable.
Quality checks of the collection immediately after it was constructed found no problems, though those checks necessarily used the same set of runs and judgments available at the time.

Twenty years later, the advent of new neural retrieval techniques, which many regard as being fundamentally different from the retrieval techniques used in the late 1990's, allows us to directly test the reusability claim for a set of new runs.
The new runs {\em did} retrieve previously unretrieved and hence unjudged documents.
These unjudged documents were examined by TREC assessors who found the vast majority of them to be not relevant.
Ordering systems by mean evaluation score computed using the original qrels on the one hand and the qrels augmented with the new judgments on the other produced practically identical ranked lists.
For these runs, we demonstrate the \mbox{TREC-8} ad~hoc collection to be a reliable measurement tool.

The fact that these runs retrieved (very) few newly found relevant documents is not proof that some other run ({\it magical}) would not retrieve more, of course.
Absolute certainty regarding the relevant set would require judging the entire corpus for each topic.
But for a new run to score materially differently from the score produced by the \mbox{TREC-8} qrels it must both retrieve a substantial number of currently unknown relevant documents and rank the newly-found relevant before the majority of the known relevant.
We believe such a result to be unlikely.

Unfortunately, ``use deep pools over highly effective runs'' is not very actionable advice for building new reusable test collections.
The document corpus used in the \mbox{TREC-8} collection is about 525,000 documents---fairly modest by today's standards---and pooling is known to have its own size dependency.
The work reported here provides neither new insights into how to build high-quality collections in the first place nor new tools for gauging collection quality.
As both of these problems are areas of active research, we leave them for future work.
But for now, we can sleep more easily using the \mbox{TREC-8} ad~hoc test collection.

\begin{acks}

We'd like the thank Xueguang Ma and Xinyu (Crystina) Zhang for generating the neural runs in our experiments.
This research was supported in part by the Natural Sciences and Engineering Research Council (NSERC) of Canada, with computational resources provided by Compute Ontario and Compute Canada.
\end{acks}

\bibliographystyle{ACM-Reference-Format}
\bibliography{neural-eval}

\end{document}